\documentclass[12pt,preprint]{aastex}

\newcommand\teff{T_{\rm eff}}

\newcommand{\e}[1]{ \times 10^{#1}}
\newcommand{\wig}[1]{\mathrel{\hbox{\hbox to 0pt{%
          \lower.6ex\hbox{$\sim$}\hss}\raise.4ex\hbox{$#1$}}}}

\shorttitle{short title}
\shortauthors{Saumon et al.}

\begin{document}


\title{New H$_2$ collision-induced absorption and NH$_3$ opacity and the
       spectra of the coolest brown dwarfs}


\author{Didier Saumon}
\affil{Los Alamos National Laboratory, PO Box 1663, Mail Stop F663, Los Alamos, NM 87545}
\email{dsaumon@lanl.gov}

\author{Mark S. Marley}
\affil{NASA Ames Research Center, Mail Stop 245-3, Moffett Field CA 94035}
\email{Mark.S.Marley@nasa.gov}

\author{Martin Abel}
\affil{Physics Department, University of Texas at Austin, Austin, TX, 78712}
\email{mabel@physics.utexas.edu}

\author{Lothar Frommhold}
\affil{Physics Department, University of Texas at Austin, Austin, TX, 78712}
\email{frommhold@physics.utexas.edu}

\author{Richard S. Freedman}
\affil{SETI Institute, 515 Whisman Rd, Mountain View CA 94043}
\affil{Space Science and Astrobiology Division, NASA Ames Research Center, Mail Stop 245-3, Moffett Field CA 94035}
\email{freedman@darkstar.arc.nasa.gov}



\begin{abstract}
We present new cloudy and cloudless model atmospheres for brown dwarfs using recent
ab initio calculations of the line list of ammonia (NH$_3$) and of the 
collision-induced absorption of molecular hydrogen (H$_2$).  We compare
the new synthetic spectra with models based on an earlier description of the H$_2$ and 
NH$_3$ opacities. We find a significant improvement in fitting
the nearly complete spectral energy distribution of the T7p dwarf Gliese 570D and
in near infrared color-magnitude diagrams of field brown dwarfs. We apply these new
models to the identification of NH$_3$ absorption in the $H$ band peak of very late
T dwarfs and the new Y dwarfs and discuss the observed trend in the NH$_3-H$ spectral index. 
The new NH$_3$ line list also allows a detailed
study of the medium resolution spectrum of the T9/T10 dwarf UGPS J072227.51$-$054031.2
where we identify several specific features caused by NH$_3$. 

\end{abstract}





\section{Introduction}

Brown dwarfs, with masses between $\sim 0.01$ -- 0.075$\,M_\odot$, are not massive enough
to maintain steady hydrogen fusion in their cores and, except when they are very young, have effective 
temperatures below that of main sequence stars.  Their low atmospheric temperatures allow 
the formation of molecules and condensates that leave conspicuous imprints on their 
spectral energy distribution (SED) that have led to the creation of two spectral classes:
the L dwarfs with $\teff \sim 2400$ -- 1400$\,$K \citep{kirk05} and the cooler T dwarfs extending
down to $\teff \sim 500\,$K \citep{leggett10a}.  The recent discovery of several extremely cool dwarfs by the
Wide-field Infrared Survey Explorer (WISE) motivates the creation of a new spectral
class: the Y dwarfs with $\teff \wig< 500\,$K \citep{cushing11}.

The spectra of L dwarfs are characterized by absorption bands of VO, TiO, H$_2$O, CO, FeH, CrH and
the very broad doublets of Na I (0.59$\,\mu$m) and K I (0.77$\,\mu$m) \citep{kirk99}.  
While condensates, primarily corundum, magnesium silicates and iron, do not have spectral
features as distinctive as those of molecules, their continuum opacity reddens the near-infrared spectrum of L dwarfs 
significantly.  The effects of the cloud opacity peak at $\teff \sim 1500$ --1600$\,$K or spectral type L6--L8.
In the cooler T dwarfs, the condensate cloud layer sinks below the photosphere, TiO and VO have 
been lost to TiO$_2$, VO$_2$ and condensates \citep{bs99,lod02} the carbon
chemistry has shifted to favor CH$_4$ \citep{burgasser02,lf02}, and NH$_3$ bands appear in the mid-infrared around T2 \citep{cushing06}.
In the very latest T dwarfs and in the new Y dwarfs, weak bands of NH$_3$ appear in the near-infrared
\citep{delorme08,cushing11}. 

Due to their additional degrees of freedom and greater complexity, the opacities of molecules are not
nearly as well understood as those of atoms and ions.  
The primary molecular opacity data bases are the HITRAN \citep{hitran03,hitran05,hitran09} and the GEISA \citep{geisa99,geisa05} 
compilations, which were developed for modeling the Earth's atmosphere and consider the opacity of gases at temperatures
of 300$\,$K and below. Since the temperature in brown dwarf atmospheres can be up to 10 times higher, those
line lists are missing the so-called ``hot bands'' that arise from the absorption of a photon from an excited
lower level that is not thermally populated at 300$\,$K. 
In some cases, the molecular line spectrum is so rich that 
individual transitions are unresolved in laboratory spectra or it becomes impossible to assign 
the observed transitions to specific molecular energy levels.  The line lists are thus incomplete
for the conditions encountered in brown dwarf atmospheres,
especially for higher photon energies, hot bands, and weak transitions corresponding to near infrared and optical wavelengths.
Incomplete molecular line lists limit the predictive capability of atmosphere
models and synthetic spectra of brown dwarfs \citep{cushing08, stephens09}, a problem that 
was recognized in the first attempts to 
model the observed spectrum of a brown dwarf, the T7p dwarf Gliese 229B \citep{marley96,allard96}.
In view of the difficulties in analyzing laboratory spectra of high temperature molecular gases,
an alternative is to compute a molecular line list with a first-principles approach.

The past 15 years of brown dwarf research have motivated the theoretical modeling of the opacity of
several key molecules. New molecular line lists for H$_2$O \citep{ps97, barber06}, CO \citep{goorvitch94}, 
TiO \citep{schwenke98}, VO \citep{plez98}, FeH \citep{dulick03}, CrH \citep{burrows02}, have improved the quality 
of the brown dwarf model spectra significantly. Methane (CH$_4$) has been and remains the most
important molecule observed in brown dwarf spectra (defining the T spectral class, but appearing as
early as L5; \citep{noll00}) with a very incomplete line list at near infrared wavelengths although
some progress has been made \citep{fml08,schwenke02}. The two most recent
developments in this area are new calculations of a line list of ammonia (NH$_3$) and of the continuum
opacity from collision-induced absorption (CIA) by H$_2$ molecules.

We study the effect of these two much anticipated new opacity calculations on the spectra and
colors of brown dwarfs. In section \S2, we discuss the nature of H$_2$ CIA opacity, how the new
calculation changes the opacity and how if affects the SED of cool brown dwarfs.  The role of the NH$_3$
opacity in brown dwarf spectra is presented in section \S3, where we show how the new
line list for NH$_3$ affects primarily the near-infrared spectrum and brings new spectral 
features that can help define the T/Y dwarf boundary. 
We demonstrate the resulting improvement in synthetic spectra by comparing with the
well-characterized SED of the T7.5 dwarf Gliese 570D and near infrared color-magnitude diagrams
in section \S4. The improved NH$_3$ line list provides the opportunity to comment on the role
of ammonia in extremely cool dwarfs near the T/Y transition and on recent detections of this molecule in the spectra of 
extreme objects (section \S5). Concluding remarks are provided in section \S6.

\section{Collision-induced absorption by H$_2$}

Molecular hydrogen H$_2$ is the most abundant component of brown dwarf atmospheres. Because of its high degree of
symmetry, the H$_2$ molecule has no permanent dipole moment and only quadrupole transitions are allowed,
which are very weak. During a collision with another particle, however, the compound H$_2$-particle system has
a temporary dipole moment and can thus undergo a radiative dipole transition.  
The collision-induced absorption process is significant only at relatively high densities 
where collisions are frequent, such as in the atmospheres of planets and brown dwarfs. 
Three-body and higher order
collisions also contribute to CIA but are probably negligible in brown dwarf atmospheres \citep{hw58}.

While the probability of a dipole transition in a typical molecule involves only molecular properties,
CIA also depends on the frequency of collisions with other atoms or molecules (primarily
other H$_2$ and He, by far the most abundant species), which is linearly dependent on the density of 
the gas. For a given metallicity, the relative importance of CIA absorption in a
brown dwarf atmosphere is a measure of the density of the gas, which is closely
tied to the surface gravity.  On the other hand, the importance of H$_2$ CIA decreases in higher
metallicity atmospheres as the absorption from molecules involving heavier elements (CH$_4$, CO, H$_2$O,
NH$_3$) becomes dominant. This sensitivity of H$_2$ CIA to gravity and metallicity is a useful
spectral diagnostic in T dwarfs \citep{burgasser02,burgasser03_sd,burgasser04b}.

The H$_2$ CIA transitions occur between vibrational-rotational levels of the H$_2$ molecule and form a 
series of bands corresponding to $\Delta \upsilon=0$, 1, 2, 3, etc., where $\upsilon$ is the vibrational quantum number
of the molecule.
Each band is formed by transitions between the different rotational levels for a 
given $\Delta \upsilon$. Collisional broadening
of the individual lines is comparable to the line spacing and thus H$_2$ CIA is effectively a
continuum source of opacity.  Borysow and her collaborators calculated the H$_2$-H$_2$ and H$_2$-He
CIA coefficients in the temperature regime of interest for brown dwarf and white dwarf atmospheres 
\citep{bf86,bfm89,jhbf00,bjf01,borysow02,gf03}\footnote{Additional H$_2$ CIA references as well as codes and tables 
can be found at {\tt http://www.astro.ku.dk/$\sim$aborysow/programs/index.html}}, and
those are universally used by brown dwarf modelers \citep{fml08,sb07,allard01}.
The H$_2$-H$_2$ and H$_2$-He CIA coefficients are shown in Figures 1 and 2, respectively. 
While these two sources of opacity have similar frequency ($\nu$) and temperature ($T$) dependences, H$_2$-He 
plays a secondary role since the collisions with He are about 10 times less frequent than with H$_2$
as follows from their relative number densities. The absorption coefficient per H$_2$ molecule is
\begin{equation}
   \kappa_{{\rm H}_2}(\nu,T)= n_{{\rm H}_2}\sigma_{{\rm H}_2-{\rm H}_2}(\nu,T) + n_{\rm He}\sigma_{{\rm H}_2-{\rm He}}(\nu,T),
\end{equation}
where $n_i$ is the number density of species $i$ and $\sigma_{{\rm H}_2-i}$ is the H$_2-i$ CIA cross section.
The successive CIA bands corresponding to $\Delta \upsilon=0, 1, ... 5$ are clearly visible in Figures 1 and 2. At higher
temperatures, weaker transition become more important and fill in the inter-band regions.  
The $\Delta \upsilon=1$ band centered at $\sim 4200\,$cm$^{-1}$ (2.4$\,\mu$m) is important in
T dwarfs as it coincides with the $K$ band flux peak, which is caused by a minimum in the combined opacities of 
H$_2$O and CH$_4$.  On the other hand, the $J$ and $H$ band flux peaks in the spectra of T dwarfs are shaped 
primarily by H$_2$O and CH$_4$ absorption. Thus, the $K/J$ and $K/H$ flux ratios reflect the relative
importance of H$_2$ CIA to the H$_2$O and CH$_4$ opacities which is affected primarily by the gravity and 
metallicity of the atmosphere.  A relatively weaker $K$ band flux indicates a higher gravity (stronger H$_2$ CIA)
or a lower metallicity (weaker H$_2$O and CH$_4$ bands), or both. Our model atmospheres tend to underestimate 
the $K$ band flux, resulting in bluer $J-K$ than is  
observed for late T dwarfs \citep{sm08}. We have long suspected that this systematic offset comes form an overestimate of the strength of 
the H$_2$ CIA in the $\Delta \upsilon=1$ band. 

\subsection{New calculation of the H$_2$ collision-induced absorption}

First-principles calculations of molecular opacities are typically conducted in two steps. First, a potential
energy surface is calculated for a range of molecular geometries representing variations in bond lengths
and bond angles. This potential surface, which is multi-dimensional for polyatomic molecules, is
obtained by solving the quantum mechanical Schr\"odinger equation for the electrons for
each given nuclear geometry using methods of quantum chemistry.  The potential surface then serves as input 
for the calculation of the
eigenstates of the molecule and the induced dipole moment surface from which allowed transitions and
their oscillator strengths can be calculated to generate a line list. For bands originating from
excited states, or for relatively high energy transition (in the optical), the final state is
well above the ground state of the molecule.  The calculation of these lines requires
an accurate knowledge of the potential energy surface far from the minimum that approximately corresponds to
the ground state energy, which involves many subtle quantum mechanical effects and requires
extensive calculations for widely separated configurations of nuclei. The calculation of higher
energy transitions and of ``hot bands'' is thus increasingly difficult.

\begin{figure}
   \plotone{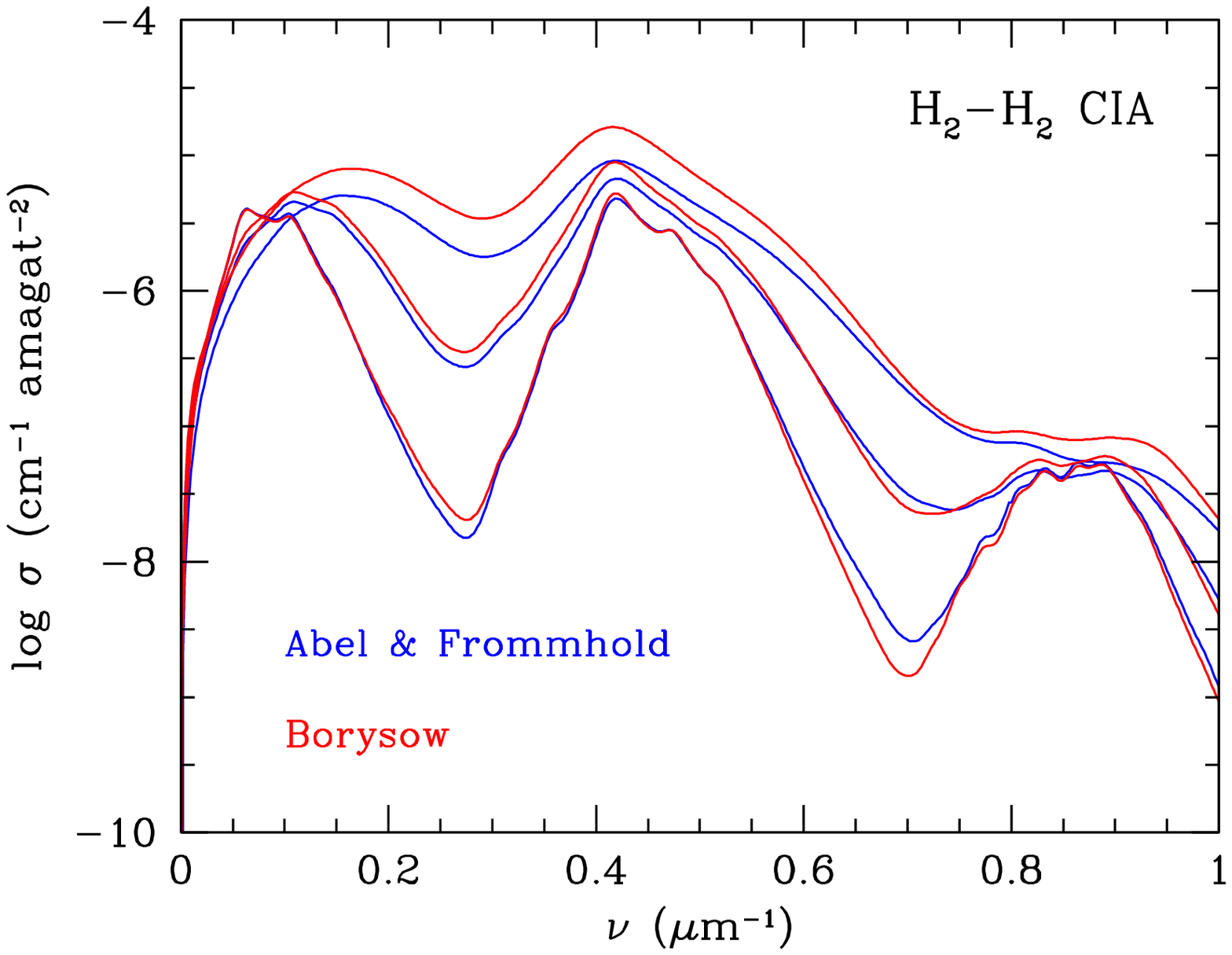}
   \caption{Collision-induced absorption coefficient for H$_2$-H$_2$ collisions at $T=500$, 1000 and 2000$\,$K, from
            bottom to top, respectively. Each broad peak corresponds to a change in the vibrational quantum number
            of $\Delta \upsilon=0$, 1, 2... from left to right, respectively. The calculations of Borysow and collaborators
            (\citet{bjf01,borysow02}, and references therein)
            and the recent work of Abel \& Frommhold \citep{frommhold10,abel11} are shown in red and blue, respectively.  
            1 amagat$=2.6867774 \e{19}\,$cm$^{-3}$.
           [{\it See the electronic edition of the Journal for a color version of this figure.}]}
    \label{cia_h2h2}
\end{figure}

\begin{figure}
   \plotone{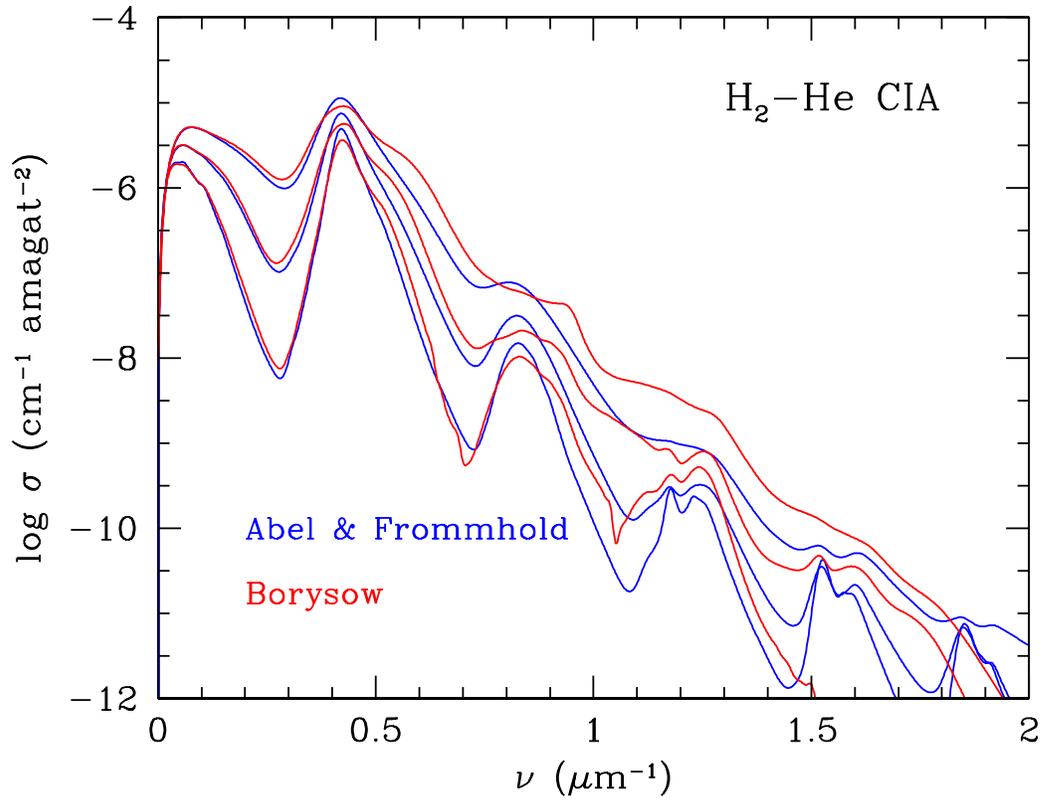}
   \caption{Same as Figure \ref{cia_h2h2} for H$_2$-He CIA. The older calculation labeled ``Borysow'' is
            from \citet{bfm89}, \citet{jhbf00}, and references therein. The ``Abel \& Frommhold''
            calculation is from \citet{abel12}.
           [{\it See the electronic edition of the Journal for a color version of this figure.}]}
    \label{cia_h2he}
\end{figure}

The growing evidence from brown dwarf spectra that something was amiss in the
$\Delta \upsilon=1$ band of H$_2$ CIA as well as applications to ultracool white dwarfs \citep{bsw95}
 prompted a new calculation of the H$_2$-H$_2$ and
H$_2$-He CIA opacity in the entire regime of temperatures and frequencies of interest \citep{frommhold10,abel11,abel12}.
New, high accuracy potential energy surfaces and induced dipole surfaces were generated for these two
CIA processes.  The resulting opacity was calculated on a fine temperature and frequency grid, from
$\nu=20$ to $10^4\,$cm$^{-1}$ and $T=200-3000\,$K for H$_2$-H$_2$ and 20 to 20000$\,$cm$^{-1}$ and
$T=200-9900\,$K for H$_2$-He \citep{richard12}. The H$_2$-H$_2$ table is being extended to the same
$(T,\nu)$ grid as the H$_2$-He table.
In each case, the entire parameter range is covered in a uniform manner, while
earlier tabulations used different approximations and potential surfaces for different temperature 
and frequency ranges.  As is the case for the older calculations, the new H$_2$-H$_2$ opacities
are in excellent agreement with experimental data, which are limited to $T \le 300\,$K \citep{frommhold_book,abel11}.
Comparisons with the CIA opacity previously used in modeling brown dwarf atmospheres are shown in 
Figures \ref{cia_h2h2} and \ref{cia_h2he}. As expected, differences are negligible or minimal at low temperatures.
The differences become substantial at 2000$\,$K in the roto-translational band for H$_2$-H$_2$ ($\sim 1000-2000\,$cm$^{-1}$)
and in all roto-vibrational bands ($\Delta \upsilon > 0$). In the fundamental band ($\Delta \upsilon=1$),
which affects the $K$ band flux in brown dwarfs, the H$_2$-H$_2$ CIA has decreased by 24\% at 1000K
and by 44\% at 2000K, which goes in the direction required to bring the $K$ band flux of the models
into better agreement with the observations.  The blue side of the fundamental band of H$_2$-H$_2$ CIA 
extends into the $H$ photometric band where the absorption has decreased somewhat at temperatures 
above $\sim 1500\,$K.  The H$_2$-He contribution (Fig. 2) shows similar variations,
except that the absorption coefficient increases at the peaks of the fundamental and first overtone bands. 
At frequencies above $10^4\,$cm$^{-1}$, the corrections can be factors of 3 or more. This matters less in brown
dwarfs as H$_2$ CIA is not an important source of opacity at optical wavelengths and the He
abundance is $\sim 10$\% that of H$_2$. 

\begin{figure}
\epsscale{0.90}
   \plotone{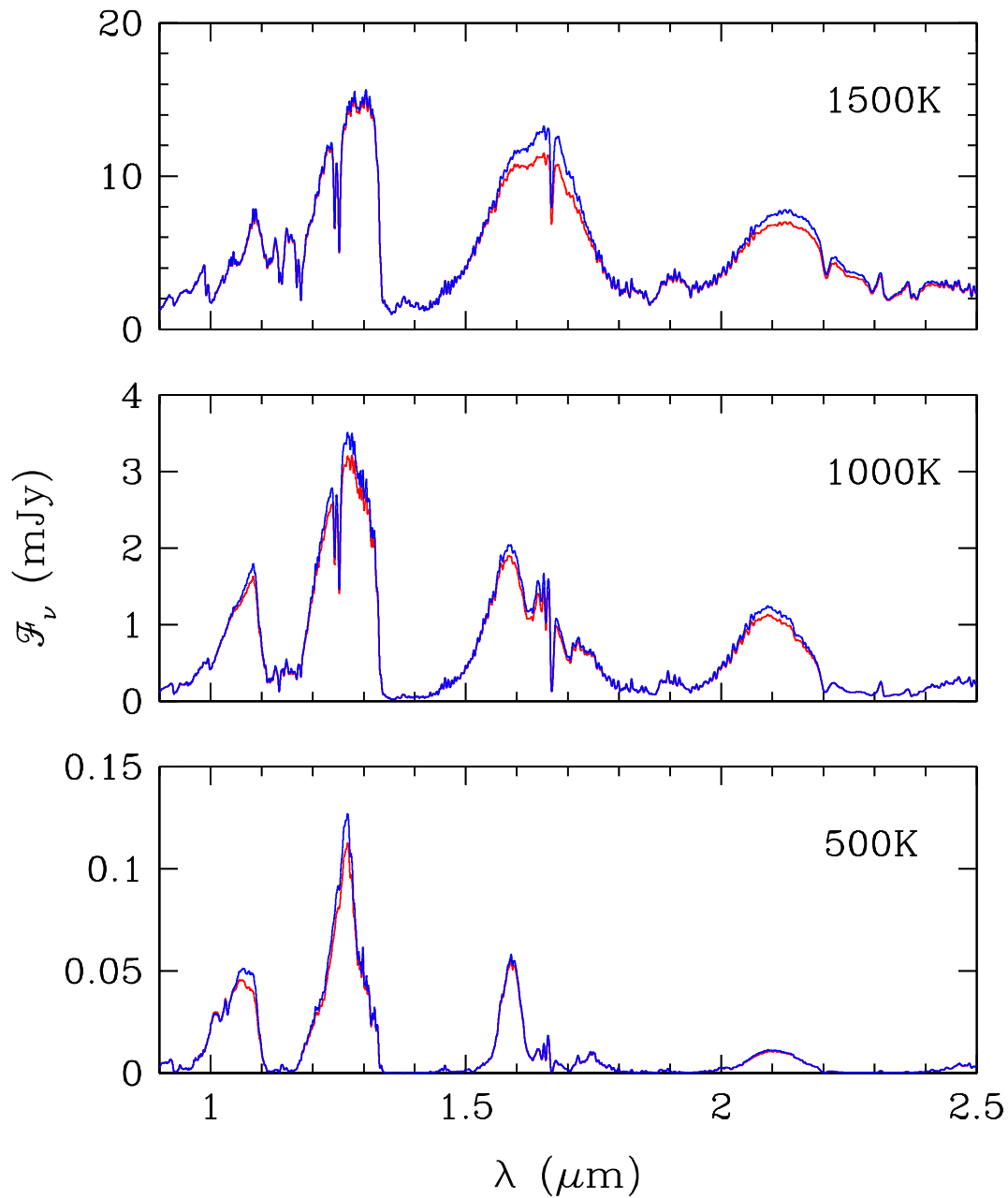}
   \caption{Effect of the new H$_2$-H$_2$ and H$_2$-He CIA opacity on synthetic spectra of brown dwarfs.  The spectra shown are
            {\it cloudless} models with $\teff=1500$, 1000 and 500$\,$K, with $\log g=5$ (cgs) and solar 
            metallicity.  The spectra computed with the new CIA opacities are shown in blue.  The red lines
            show spectra computed with the older CIA opacity and the same $(T,P)$ structures. The fluxes are calculated for
             $d=10\,$pc and are displayed at a resolving power $R=500$.}
    \label{spectra_cia}
\end{figure}

The effect of the change in the CIA opacity is shown in Fig. \ref{spectra_cia}
where spectra computed with the old and the new opacity but with the same $(T,P)$ atmospheric structures,
are compared.  All the flux peaks in the near infrared are affected, with rather modest flux increases of
$\wig< 15\%$. Since the new H$_2$ CIA opacity is nearly identical to the older calculation
at $T=300\,$K and increasingly differs at higher temperatures, 
the differences in the spectra tend to be larger at higher $\teff$ and nearly vanish for 
$\teff\wig<500\,$K except in the $Y$ and $J$ bands which are relatively transparent and probe the
atmosphere at temperatures well above $\teff$. Since the CIA opacity increases with the gas density,
i.e. with gravity, lower gravity spectra are less affected by the new CIA. For example,
at $\log g=4$ (cgs), the flux in the peaks of the SED increases by less than 7\%. At effective 
temperatures higher than those shown in Fig. \ref{spectra_cia}, a greater change in
the spectrum would be expected {\it in cloudless models}. However, in brown dwarfs atmospheres 
with $\teff\wig>1400\,$K the continuum opacity of the condensates overwhelms the H$_2$ CIA opacity and
the spectrum is unaffected.

\section{Ammonia opacity}

In brown dwarf atmospheres, the chemistry of nitrogen is dominated by molecular nitrogen N$_2$ and
ammonia NH$_3$, the latter being favored at higher pressures and lower temperatures \citep{lf02}.
The transformation of N$_2$ into NH$_3$ occurs at $\teff\sim 800\,$K and is, along with the similar 
conversion of CO to CH$_4$, one of the most significant chemical changes
in brown dwarf atmospheres as $\teff$ decreases.  While N$_2$ has
no permanent dipole moment and is effectively invisible in brown dwarf spectra\footnote{Like H$_2$, N$_2$
causes collision-induced absorption \citep{bf86_n2,bf87}, but its mole fraction of $\sim 10^{-4.2}$ makes this
contribution to the opacity in brown dwarf atmospheres completely negligible.} the fundamental band of NH$_3$ 
in the 10--11$\,\mu$m region is seen in T2 and later dwarfs \citep{cushing06}
and the entire 9--14$\,\mu$m region is affected by NH$_3$ absorption \citep{saumon06}.
Ammonia has a series of weaker overtone bands in the near infrared, two of which are centered at 2.0 and 1.5$\,\mu$m
and fall in the $K$ and $H$ photometric bands where the SEDs of T dwarfs show emission. The apparition of these
bands in the spectra of very cool brown dwarfs has been suggested as a trigger for a new spectral class
\citep{bsl03,leggett07y,kirk08}.

\citet{saumon00} reported the detection of three NH$_3$ features in a $R=3100$ spectrum of the 
T7p Gliese 229B in the 2.02 -- 2.045$\,\mu$m range but other expected NH$_3$ features were missing.
This tentative detection was followed by a decade where no other near-infrared spectrum of T dwarfs 
was taken at high enough resolution in the $K$ band to reveal NH$_3$ features.  The first unambiguous 
detection of NH$_3$ in brown dwarfs was achieved with the Infrared Spectrograph of
the {\it Spitzer} space telescope
that clearly showed the 10--11$\,\mu$m band in the combined spectrum of $\epsilon$ Indi Bab (T1+T6) 
\citep{roellig04,mainzer07}
and then in all dwarfs of type T2 and later observed with the IRS \citep{cushing06,saumon07,cushing08,stephens09,leggett10b}. 
Recently, the search for extremely cool
dwarfs has uncovered several objects with $\teff \wig< 600\,$K where the blue wing of the $H$ band flux peak
appears depressed due the presence of NH$_3$ \citep{delorme08,lucas10,liu11,cushing11}.  Finally, a new 
$R \sim 6000$ near-infrared spectrum of the $\teff \sim 500\,$K T9 dwarf UGPS J072227.51$-$054031.2 contains NH$_3$
features \citep{bochanski11}. The detection of NH$_3$ in the near-infrared spectrum of several ultracool dwarfs is one of the
factors that motivate the creation of the new Y spectral class \citep{cushing11}.

The analysis of the spectral signature of NH$_3$ in T dwarfs is 
complicated by the presence of vertical mixing in the atmosphere and by the invisibility of N$_2$.
Because the time scale for the conversion of N$_2$ to NH$_3$ is very long at low temperatures,
gas eddies from very deep in the atmospheres where N$_2$ and NH$_3$ are in chemical equilibrium at high temperatures can
be transported vertically by convection and slow mixing in the radiative zone to regions of lower temperatures on 
a time scale that is much shorter 
than the N$_2 \rightarrow$ NH$_3$
conversion can take place, resulting in a reduced NH$_3$ abundance \citep{fl94, fl96, lf02, saumon06, hb07}.
This  process can reduce the observed abundance of NH$_3$ by factors of $5-10$, weakening the NH$_3$ bands noticeably.
Reduced NH$_3$ absorption in the 9$-$14$\,\mu$m region has been found in every T dwarf observed with the {\it Spitzer} IRS
with a high enough S/N ratio \citep{saumon06,mainzer07,saumon07,burgasser08b,leggett09,leggett10b} and can be fully 
explained by considering the convective mixing time scale and
the time scale for N$_2 \rightarrow$ NH$_3$ conversion.  Considering the relatively well understood chemistry of nitrogen
and the universality of a deep convection zone in the atmospheres of brown dwarfs, it is reasonable to assume that
the abundance of NH$_3$ is reduced by vertical transport in all T dwarfs.

In view of the currently very limited opportunities for spectroscopic observations of brown dwarfs in the mid-infrared and
the relative ease of near-infrared spectroscopy,
the classification of the growing number of very late T dwarfs and of the new Y dwarfs relies primarily on their
near infrared spectral characteristics where NH$_3$ figures prominently as a new spectral index indicator 
\citep{delorme08,lucas10,cushing11,burgasser11} as anticipated by \citet{bsl03,leggett07y} and
\citet{kirk08}.  The importance of NH$_3$ as a strong absorber in the
mid-infrared, an indicator of non-equilibrium chemistry, and an important marker for the spectral classification of 
very late T dwarfs and Y dwarfs demands a high quality line list to compute reliable models.

\subsection{New theoretical NH$_3$ line list}

\begin{figure}
\epsscale{1.00}
   \plotone{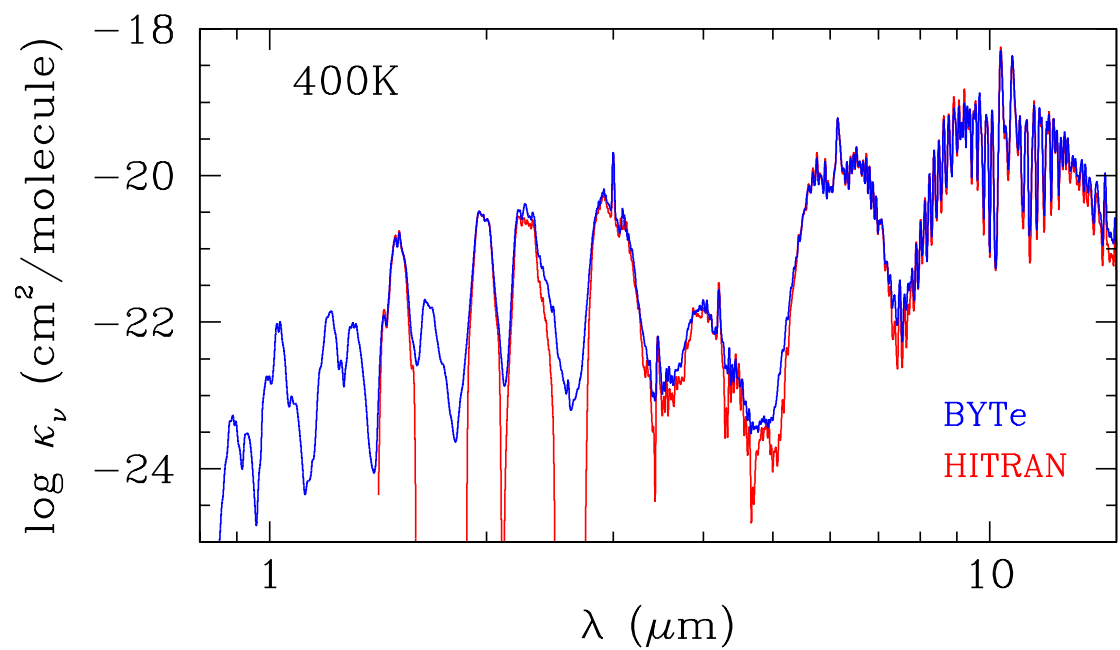}
   \caption{Absorption coefficient of NH$_3$ at $T=400\,$K and $P=1\,$bar.  The red line shows
            the opacity calculated from the line list that we have used in our previous atmosphere models. It is
            primarily based on the HITRAN data base supplemented with more recent laboratory
            measurements \citep{fml08}. The blue line labeled ``BYTe'' shows the absorption coefficient
            computed from the new first-principles line list of \citet{BYTe}. 
            For clarity, the absorption coefficient is shown at a resolving power of $R=200$.}
    \label{nh3_400}
\end{figure}

\begin{figure}
\epsscale{1.00}
   \plotone{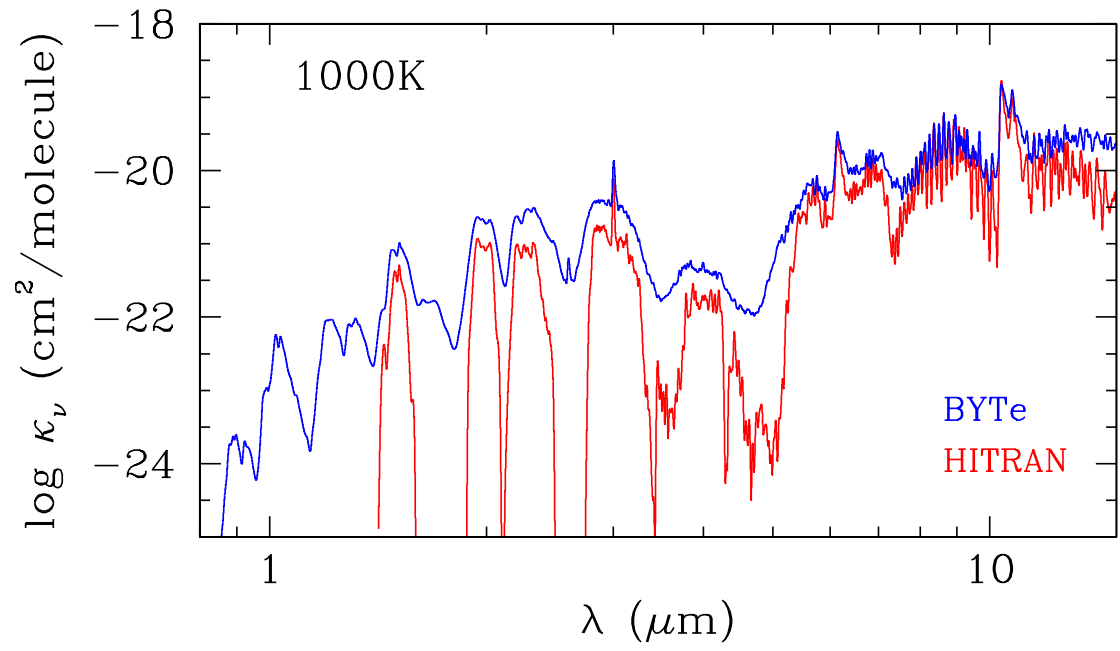}
   \caption{Same as Figure \ref{nh3_400} for $T=1000\,$K.}
    \label{nh3_1000}
\end{figure}

Our previous models used a NH$_3$ line list from the HITRAN data base \citep{hitran03,hitran05,hitran09} supplemented with room 
temperature data in the 6000$-$7000$\,$cm$^{-1}$ range \citep{fml08,sb07}.
In principle, the fundamental band at $10-11\,\mu$m should be well represented in the low temperature ($\sim 300\,$K)
HITRAN data base but the line list becomes increasingly incomplete at shorter wavelengths and higher temperatures. 
The resulting opacity at 400$\,$ and 1000$\,$K  is shown in Figs. \ref{nh3_400} and \ref{nh3_1000}. There are several gaps
in the inter band regions 2.5--2.75$\,\mu$m, $\sim 2.11\,\mu$m and 1.6--1.88$\,\mu$m, and the list stops 
at 1.42$\,\mu$m. Such a line list is inadequate to analyze the rapidly growing spectroscopic data for
$\teff \wig< 600\,$K dwarfs, which has motivated recent experimental \citep{hargreaves11} and theoretical work
on ammonia \citep{huang11a,huang11b,yur11}.

A new extensive line list for NH$_3$ has recently been calculated from first principles.  It is based on a new
potential energy surface \citep{yur05} that has been refined by fitting calculated energy levels to
nearly 400 levels that are accurately known experimentally \citep{yur11}.  The resulting energy levels
agree with the experimental values to within 1$\,$cm$^{-1}$ and generally under 0.1$\,$cm$^{-1}$.
This `BYTe' NH$_3$ line list \citep{BYTe} contains over 1.1 billion lines, compared to $\sim 34000$ in HITRAN. 
It extends from 0 to 12000$\,$cm$^{-1}$ (0.83$\,\mu$m) and is applicable up to 1500$\,$K. An independent 
calculation of a NH$_3$ line list is under way \citep{huang11b}.  Figures \ref{nh3_400} and
\ref{nh3_1000} show the improvement of the BYTe line list over the HITRAN opacity. At 400$\,$K, we expect the HITRAN opacity to
be quite good since it is based on data acquired primarily at room temperature.  This is indeed what the BYTe
line list shows as both are nearly identical near the center of all the bands in common between the two line lists. 
In addition, the BYTe line list includes new bands at 1.65$\,\mu$m and
shortward of 1.4$\,\mu$m, and contains many weak lines that fill in the inter band regions where there are gaps in
the HITRAN line list.  The differences become more dramatic at higher temperatures (Fig. \ref{nh3_1000}) where the 
transitions from thermally excited levels come into play, reducing the contrast between the peaks and valleys of the 
opacity and increasing the band
opacity by factors of $\sim 3-5$ compared to the HITRAN value. The opacity in the $8-15\,\mu$m also increases significantly.

\begin{figure}
\epsscale{0.90}
   \plotone{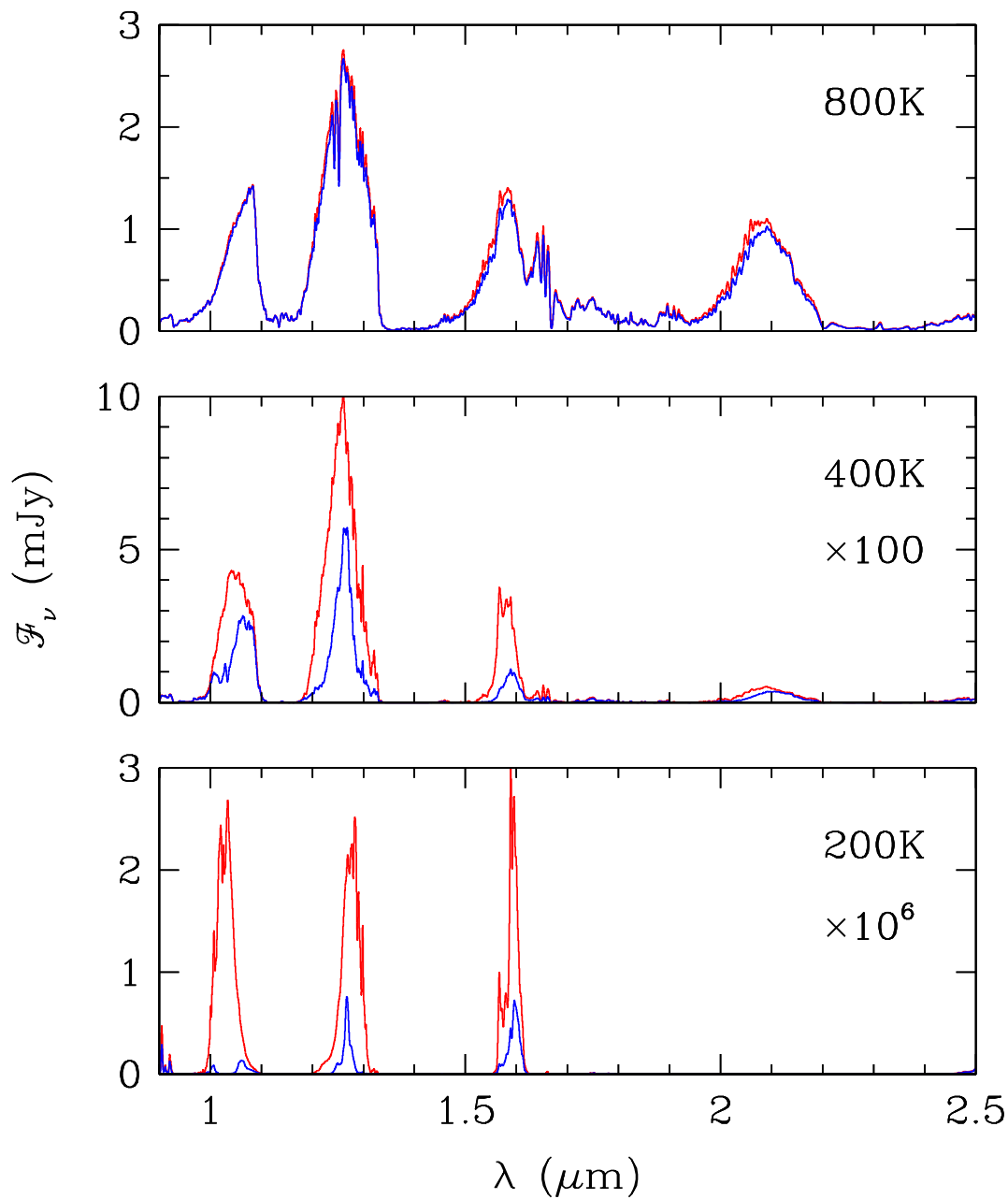}
   \caption{Effect of the new NH$_3$ opacity on near infrared spectra of cool brown dwarfs. Three models
            are shown with $\teff=800$, 400 and 200$\,$K.  All models
            shown have solar composition, $\log g=4$, are cloudless and in chemical equilibrium. Models 
            computed with the HITRAN line list for NH$_3$ are shown in red while those computed from
            the same $(T,P)$ structure but with the BYTe NH$_3$ line list are in blue.
            The spectra are shown as absolute fluxes for a distance of 10$\,$pc and at $R=500$ for clarity.}
    \label{spectra_nh3}
\end{figure}

The effect of the change in NH$_3$ opacity on the synthetic spectra is shown in Fig. \ref{spectra_nh3}. For this comparison,
spectra computed with the HITRAN and the BYTe line lists use the same $(T,P)$ atmosphere structure. At $\teff=1200\,$K (not shown),
the spectra are identical but at $\teff=800\,$K there is a small
flux decrease of $\wig< 20$\% on the blue side of both the $H$ and $K$ peaks.
At lower $\teff$, the chemical equilibrium shifts to strongly favor NH$_3$ over N$_2$ as the dominant nitrogen bearing
species and its opacity becomes more significant compared to the competing absorption from H$_2$O and CH$_4$. This leads to
dramatic changes in the spectrum. All the near infrared flux peaks are reduced by factors of
$2-4$. In particular, the new bands at short wavelengths dramatically affect the shape of the $Y$ and $J$ peaks at $\teff=400\,$K.
At the extremely low $\teff$ of 200$\,$K, the changes are limited to the spectral regions were there are gaps in the HITRAN line list,
and lead to very distinctive features in the $Y$ and $J$ bands.  In particular, the $Y$ band peak is split in two.
Figure \ref{spectra_nh3} suggests that in addition to the
NH$_3-H$ spectral index which more or less measures the slope of the blue side of the $H$ band flux peak
\citep{delorme08}, another NH$_3$-sensitive index could be defined to measure the curvature of the $J$ band peak.
In the mid-infrared (not shown), the most variation is shown by the 800$\,$K spectrum with a $<40$\% decrease in flux in 
the $9-15\,\mu$m region.  The 400 and 200$\,$K spectra are barely affected in the mid-infrared.
All these differences decrease somewhat as gravity is increased. The spectra shown in Fig. \ref{spectra_nh3} and discussed
above are computed in chemical equilibrium. However, we expect that the NH$_3$ abundance will be prevented from coming into
equilibrium by convection, resulting in surface abundances $\sim 10$ times lower in the upper atmosphere.  Consequently, 
NH$_3$ absorption is more muted in non-equilibrium spectra and the changes discussed above are smaller than in
model that are in chemical equilibrium.  The one exception is the $\teff=200\,$K model which is so cold that
the entire atmosphere is dominated by NH$_3$ whose abundance becomes nearly insensitive to convective mixing. 

The recently identified Y dwarfs have $\teff \sim 300 - 500\,$K \citep{cushing11}. The spectra computed with the new BYTe
line list indicate that broad NH$_3$ features should be detectable in the near infrared, most notably as a shoulder on
the blue side of the $Y$ band peak at $\teff \sim 400\,$K which is split at $\teff\wig<300\,$K by a growing NH$_3$ band 
centered at 1.03$\,\mu$m. The $J$ band peak becomes triangular and the $H$ band peak narrows.  The strength of these
effects is reduced by the non-equilibrium abundance of NH$_3$.

\section{Models including both the new H$_2$ CIA and NH$_3$ opacities}

\begin{figure}
\epsscale{0.85}
   \plotone{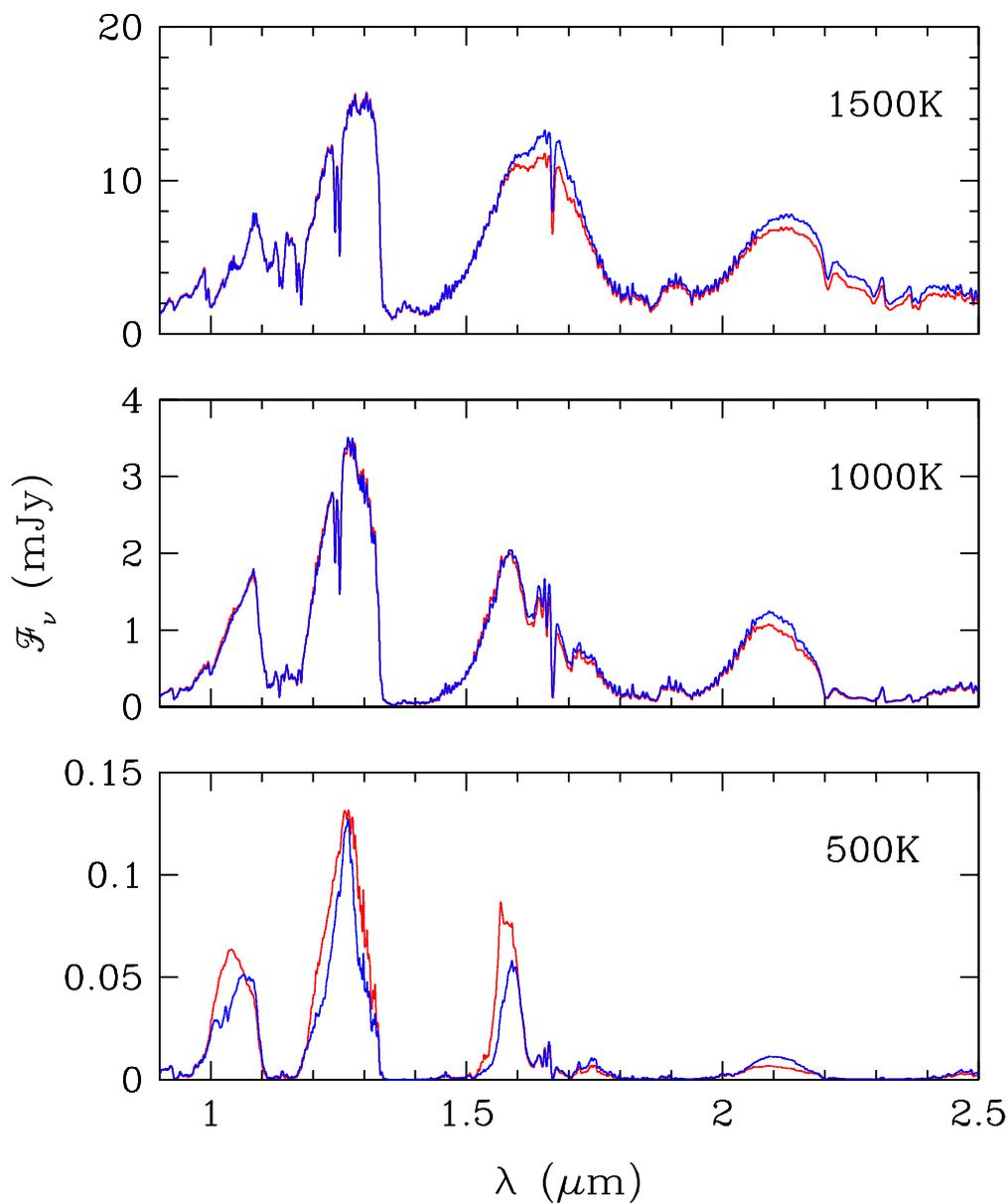}
   \caption{Comparison of spectra including the new H$_2$ CIA and the BYTe NH$_3$ opacities (blue) with spectra from
            a previous
            generation that only differ in these two sources of opacity (red).  These cloudless spectra 
            have $\log g=5$ (cgs), solar metallicity and are computed in chemical equilibrium.  The
            effective temperatures are 1500, 1000 and 500$\,$K.
            In contrast with Figs. \ref{spectra_cia} and \ref{spectra_nh3}, the atmospheric $(T,P)$ structures
            have been computed with the appropriate (old or new) opacities.
            The spectra are plotted at $R=500$ for clarity.}
    \label{spectra}
\end{figure}

We have computed new model atmospheres and synthetic spectra using the new H$_2$ CIA and NH$_3$ opacities.  In all other
aspects, these models are identical to those described in \citet{sm08,mar02} with additional details provided in \citet{cushing08}
and \citet{stephens09}. These new models, which include cloudless and cloudy atmospheres, were used in the recent analyzes 
presented in \citet{cushing11,luhman11} and \citet{leggett11}.
These new cloudless model spectra are compared to our previous models in 
Fig. \ref{spectra}, where, in contrast with Figs. \ref{spectra_cia} and \ref{spectra_nh3},
the $(T,P)$ structures are computed with their respective opacities and thus the model structures and spectra are self-consistent. In this case,
the flux is redistributed along the SED to conserve the total flux (which was not the case in Figs. \ref{spectra_cia} and \ref{spectra_nh3}). 
At $\teff=1500$ and 1000$\,$K, the changes are modest and dominated by the new H$_2$ CIA opacity.  The  reduced opacity in the
first overtone band of H$_2$-H$_2$ CIA is responsible for the increased flux in the $K$ band peak ($\sim 15$\%) and a smaller increase in the
$H$ band flux.  The $\teff=1500\,$K spectrum also shows a 7\% flux increase in the 4$\,\mu$m peak but the 3--15$\,\mu$m cloudless spectra
are otherwise unaffected. Near the T/Y dwarf transition, where $\teff \sim 500\,$K, all four near infrared peaks in the spectrum are noticeably
affected. Absorption by NH$_3$ changes the shape of the $Y$ and $J$ peaks and shaves off the blue side of the $H$ band peak.  Back warming
from this NH$_3$ absorption pushes the flux out in the $K$ band, which is now $\sim 75$\% brighter, and in the 4--8$\,\mu$m
region where the flux increases by $\sim 30$\%.

\subsection{Comparison with the spectrum of Gliese 570D}

The T7.5 dwarf Gliese 570D \citep{burgasser00} is the most thoroughly studied late T dwarf. Optical, 
near infrared, $M$ band, and {\it Spitzer} IRS spectroscopy
are available -- sampling $\sim 70$\% of the total flux emitted -- as well as {\it Spitzer} IRAC and WISE photometry.  
As this brown dwarf is a companion to a nearby main sequence star, its distance and
metallicity are accurately known and its age is fairly well constrained to 2--5$\,$Gyr. Because Gl 570D is nearby and relatively bright, the whole data
set is of high quality and very effectively constrains the models.  It is an excellent test object to validate the new models and evaluate the improvements 
brought about by the new opacities.  Detailed analyzes of Gl 570D \citep{geballe01,saumon06,geballe09}
have established that $\teff=800-820\,$K and $\log g=5.09-5.23$ (cgs), where the ranges of $\teff$ and $\log g$ are correlated: 
a higher $\teff$ corresponds 
to a higher gravity. Solar metallicity was adopted based on the metallicity of the primary ([Fe/H]$=0.09 \pm 0.04$). \citet{saumon06} 
established that convective transport drives the NH$_3$ abundance below the value expected from chemical equilibrium and
obtained an very good fit of the entire SED of Gl 570D (see also \citet{geballe09}). Figure \ref{gl570d} shows the spectrum
computed by \citet{saumon06} for those parameters within the above range that best reproduce the data
and a new model computed with the same atmospheric parameters but with the new H$_2$ CIA and NH$_3$ opacities. There is
significant improvement in the $K$ band peak which is now fitted very well. There is also a modest improvement in
the 5.5--7.5$\,\mu$m region and beyond 9$\,\mu$m where NH$_3$ absorption features overlap with H$_2$O absorption. The new
H$_2$ CIA calculation, which is primarily responsible for the increased flux in the $K$ band represents a significant
improvement over the previously available tabulation. Outside of the 9--14$\,\mu$m region, NH$_3$ absorption is rather weak at this $\teff$,
and the new NH$_3$ opacity leads to very modest changes.  
We believe that the remaining discrepancies in our fot of Gl 570D, such as the
width of the $J$ band peak and the depth of the 1.6$\,\mu$m band of CH$_4$, are primarily due to the very incomplete methane
line list, which is presently in a similar state as the ammonia line list was prior to the first-principles calculation
of \citet{BYTe}.

\subsection{Comparison with near-infrared photometry}

While a comparison with the SED of Gl 570D provides a detailed test of the new models, the overall effect of the new opacities
can be better appreciated in color-magnitude diagrams where synthetic colors can be compared to a large set of photometric data.
Prior to 2003, our models used the solar abundances of 
\citet{lf98} and the cloudless sequence was in generally good agreement with the near infrared colors of late T dwarfs \citep{burgasser02}.
A significant downward revision of the CNO abundances in the solar photosphere \citep{allende02} motivated a change of the
solar abundances in our models \citep{lod03} which moved the cloudless sequence to the blue in $J-K$.
The resulting sequence of [M/H]=0 cloudless models are too blue compared to late field T dwarfs, particularly in 
$J-K$, when compared to the near infrared colors of late-T field dwarfs (Fig. \ref{ccd1} and Fig. 10e of \citet{sm08})
\footnote{Note that the cloudless COND models of \citet{allard01} use the older solar abundances of \citet{gn93} and agree
very well with the observed near-infrared colors of late T dwarfs.}.  
The new H$_2$-H$_2$ CIA has a weaker first overtone band which increases the $K$ band flux and moves the cloudless models to 
redder $J-K$, largely compensating for the effect of the reduced CNO abundances (Fig. \ref{ccd1}). The new sequence of models
is still slightly bluer by $\wig<0.15$ than the late T dwarf sequence which may be due to missing CH$_4$ opacity on the $J$ band peak.
The spread of the observed sequence of late T dwarf colors is significant and can be accounted for by the observed distribution in
metallicity of field dwarfs (Fig. 15f of \citet{sm08}).  The very latest T dwarfs recently discovered, with $M_K\wig>17$
reveal a saturation in $J-K$ and are redder than the calculated cloudless sequences. This
may be caused by several limitations in our models
such as the near infrared opacity of CH$_4$, the presence of clouds of compounds with low condensation temperature
such as Na$_2$S, or the appearance of water clouds. Note that water condensation is included in the chemistry of the
atmosphere and the gas phase H$_2$O abundance is reduced in the low-$\teff$ models where water condenses. We are investigating 
the role of clouds in low-$\teff$ atmospheres and the results will be reported in a separate publication.  On the other hand, 
our cloudy sequences account for iron and silicate clouds and reproduce the SEDs and infrared colors of field L dwarfs fairly well
\citep{sm08,cushing08,stephens09}.  Cloudy models with $\teff\wig>900\,$K are representative of field L dwarfs and the HR 8799 planets
\citep{mar12} and are shown in Fig. \ref{ccd1}.  These hotter, cloudy models are barely affected by the new opacities and their
near infrared colors are very nearly identical to the older ones
down to $\teff=900\,$K.  This is a consequence of the opacity of condensates which reduces the importance of the H$_2$ CIA
and of the higher temperatures of the cloudy models which reduce the abundance of NH$_3$.

The same general picture emerges from the $M_J$ versus $J-H$ color-magnitude diagram (Fig. \ref{ccd2}) although the new cloudless
model now represents the $J-H$ sequence of the late field T dwarfs quite well\footnote{This may be somewhat fortuitous in view of the model
limitations mentioned above, particularly the poorly modeled 1.6$\,\mu$m band of CH$_4$ which affects the $H$ band 
(Fig. \ref{gl570d})}. The data suggest that
the $J-H$ color saturates around $-0.6$ (see also \citet{kirk11} at $M_J\sim17$ which is further supported by the position of the Y0 dwarf
WISEP J$1541-2250$ in this color-magnitude diagram ($M_J=23.9\pm0.8$, $J-H=0.17\pm0.63$, \citet{cushing11,kirk11}).  
Our cloudless sequences do show such a minimum in $J-H$ at $\teff \sim 400\,$K and a rapid turnover to the red at lower $\teff$. 
However, the models do not turn over to the red soon enough and reach values of $J-H \sim -1$, which is bluer than is observed.
It will be interesting to see how the near infrared color-magnitude diagrams will be affected by a more complete CH$_4$ line list
when it becomes available, or by the inclusion of a water cloud in the models.

\section{Ammonia and Y dwarfs}

Increased absorption in the near infrared has been observed on the blue side of the $H$ band
peak in the very late T dwarfs and in the new Y dwarfs and it is thought to be due to NH$_3$. The strength of 
the 1.5$\,\mu$m band of ammonia can be quantified with the NH$_3-H$ index defined by \citet{delorme08}
\begin{equation}
{\rm NH}_3-H={\int_{1.53}^{1.56} f(\lambda)\,d\lambda \over \int_{1.57}^{1.60} f(\lambda)\,d\lambda}\,.
\end{equation}
Smaller values of NH$_3-H$ indicate a relatively low flux on the blue side of the $H$ band
peak that is expected from stronger NH$_3$ absorption. A compilation of
the data for the latest dwarfs known is shown in Fig. \ref{nh3_index}. The NH$_3-H$ index strongly correlates
with the width of the $J$ band peak (measured by the $W_J$ index, \citet{warren07}), and
with spectral type \citep{burningham08,cushing11}, with later types having smaller NH$_3-H$. 
Figure \ref{nh3_index} shows a break in the slope at $W_J \sim 0.3$ corresponding to spectral type T8
\citep{cushing11}.
Based on the appearance of the blue side of the $H$ band peak of the Y0 dwarf WISEP J1738+2732 -- one
of the most extreme objects in Fig. \ref{nh3_index} -- compared to
earlier T8 and T9 dwarfs, \citet{cushing11} tentatively ascribe the absorption to NH$_3$. 

The models shown in Fig. \ref{nh3_index} reproduce the observed trend very well and 
indicate that the reduced abundance of NH$_3$ caused by non-equilibrium chemistry is favored. 
The dashed lines show the indices computed from the same models after the opacity of NH$_3$ has been removed 
with a trend that
is strongly at odds with the data. The decrease in the modeled NH$_3-H$ at $\teff \wig< 600\,$K (or spectral type later 
than T8) is caused by the 1.5$\,\mu$m band of NH$_3$.
The expectation that NH$_3$ bands would appear in the near infrared at low enough $\teff$ and its
detection as a depression in the blue side of the $H$ band peak is thus well supported on a quantitative basis by the models.
However, the non-equilibrium  models still show a systematic offset from the data, even for the low gravity of 
$\log g=4$ (cgs) which, for these low $\teff$, would correspond to a rather low mass of $\sim 10\,$M$_{\rm Jupiter}$. 
It is very likely that the incomplete CH$_4$ line list is responsible for this systematic shift 
since methane is a strong absorber in the $H$ band and a moderate absorber in the $J$ band
(Fig. \ref{gl570d}).

Without the benefit of models, or with models of limited fidelity, the
interpretation of the observed trends in spectral indices can be ambiguous,
particularly when considering their variations over a wide range of $\teff$. On the other hand, the direct 
detection of spectral features provides much more secure evidence of the presence of specific molecular species.
While the detection of NH$_3$ in the 2.02--2.045$\,\mu$m region of the spectrum of the T7p Gl 229B \citep{saumon00}
was tentative, \citet{bochanski11} identified 11 ammonia features in their medium resolution 
($R\sim 6000$) near infrared spectrum of the T9 dwarf UGPS 072227.51$-$054031.2 (here after UGPS 0722$-$05). 
Figure \ref{0722_H} shows the portion of their spectrum with
the most NH$_3$ features.  For comparison, the top curve shows the log of
the NH$_3$ opacity computed from the BYTe line list at the conditions corresponding to the atmospheric level
where this part of the spectrum is emitted.  The lowest curve is an unscaled model spectrum  
with $\teff=500\,$K and $\log g=4.25$, solar metallicity, and $K_{zz}=10^4\,$cm$^2$/s \citep{leggett11}.
The six NH$_3$ absorption features identified by \citet{bochanski11} in this part of the spectrum  all have
corresponding troughs in the model spectrum.  Four of these features also match peaks in the NH$_3$
opacity but the other two (1.5264 and 1.5411$\,\mu$m) do not and
their attribution to NH$_3$ is not secure.  
 
To better identify NH$_3$ absorption features in the spectrum of UGPS 0722$-$05 we show in Fig. \ref{0722_H}
a spectrum computed from the same atmosphere model but without any NH$_3$ opacity (blue curve).  There
is obviously much NH$_3$ absorption over the entire range shown, even considering the fact that the
model spectrum has a reduced (non-equilibrium) abundance of NH$_3$.  This is fully consistent
with the smaller value of the NH$_3-H$ index of UGPS 0722$-$05 compared to earlier T dwarfs (Fig. \ref{nh3_index}).
However,  a careful comparison of the spectra with and without NH$_3$ does not reveal any distinct
feature at this spectral resolution. Every feature in one spectrum appears in the other, albeit with
a different amplitude.  For example, the 1.5142$\,\mu$m absorption feature matches a peak in the NH$_3$ opacity and
an edge in the model spectrum (Fig. \ref{0722_H}), but that edge also appears in the spectrum computed without 
NH$_3$, making the attribution doubtful. The model spectrum reproduces individual features of the observed spectrum
fairly well (after scaling), but not exactly, which makes the attribution of specific features
problematic.  We find this to be also the case in other regions where \citet{bochanski11} found NH$_3$
features (1.21--1.27$\,\mu$m, 1.56--1.64$\,\mu$m and at 1.990$\,\mu$m).\footnote{For completeness, we note that the
two features in the low-resolution ($R\sim500$) spectrum of UGPS 0722$-$05 pointed out by \citet{lucas10} at
$\sim 1.275\,\mu$m and at $\sim 1.282\,\mu$m can be identified with synthetic spectra at higher
resolution as blends of H$_2$O features. A weaker blend of H$_2$S features also contributes to the latter.} 

The situation is different in the 1.96--2.05$\,\mu$m spectral region 
where \citet{saumon00} found 3 ammonia
features in Gl 229B. Figure \ref{0722_K} shows this spectral region with the same models and opacity shown
in Fig. \ref{0722_H} (with different scalings for clarity). The vertical solid and dashed lines correspond to
19 NH$_3$ absorption features that are clearly identifiable from the difference between the
spectrum computed with the non-equilibrium NH$_3$ abundance (red) and the spectrum computed without NH$_3$
opacity (blue).  In this spectral range, almost every peak in the NH$_3$ opacity (green) corresponds to
a distinct absorption feature in the synthetic spectrum.  The solid lines identify 8 absorption features that are present in the
spectrum of UGPS 0722-05\footnote{If we ignore the single high-valued pixel making up the spike in the data at 2.0175$\,\mu$m, 
the NH$_3$ feature stands out.}, and the dashed lines indicate 6 NH$_3$ features that are not clearly
detected in the spectrum and 5 that appear to be missing. Two of the detected features (2.0374 and 2.0455$\,\mu$m)
match those found in Gl 229B by \citet{saumon00}.  An absorption feature clearly identified in
\citet{bochanski11}  at 1.9905$\,\mu$m (dotted line) corresponds to a well-defined absorption
feature in the model spectrum. This feature is barely affected by the removal of all NH$_3$ opacity, however,
and does not match very well with the NH$_3$ opacity peak centered at 1.9896$\,\mu$m. In our model spectrum,
this absorption is caused by H$_2$O.

We conclude that the detection of ammonia absorption features in the near infrared spectra of very 
late T dwarfs and Y dwarfs is 
easier in the $K$ band near 2$\,\mu$m than in the $H$ band near 1.53$\,\mu$m. The main reason for this is
that the NH$_3$ opacity in the $K$ band is composed of a series of more or less equally spaced and well separated
peaks (Fig. \ref{0722_K}) while in the $H$ band it has a denser irregular structure with a smaller dynamic range as well as
a peak opacity that is about one order of magnitude smaller (Figs. \ref{0722_H} and \ref{nh3_400}).  With higher resolving 
power it becomes possible to isolate NH$_3$ features in the $H$ band. For example, with $R\sim20000$, about
a dozen features stand out in the 1.51--1.57$\,\mu$m region of the model shown in Fig. \ref{0722_H}. For dwarfs
cooler than $\sim 400\,$K, the condensation of water becomes significant \citep{bsl03}, removing H$_2$O from the gas
and unmasking the NH$_3$ bands.  For instance, the blue side of the $H$ band peak
becomes completely dominated by NH$_3$ absorption and NH$_3$ features should be readily detectable in the spectrum.
This is also true in the 2 -- 2.05$\,\mu$m region, although the flux is much lower than at 1.55$\,\mu$m.
The NH$_3$ bands will thus become stronger with lower $\teff$ until ammonia itself begins to condense at
$\teff\sim160\,$K \citep{bsl03}.

\section{Conclusion}

We have computed new cloudy and cloudless models and synthetic spectra\footnote{The synthetic spectra are available upon
request to D. Saumon (dsaumon@lanl.gov).} of brown dwarf atmospheres using 
recent first-principles calculations of the opacity of NH$_3$ and of the collision-induced absorption of H$_2$.
Both constitute very significant theoretical improvements in the modeling of the opacity
and, in the case of NH$_3$, the BYTe line list greatly expands over the HITRAN line list that is based primarily on
experimental results. The cloudy models, relevant to
L dwarfs, are barely affected by the new opacity because of their higher $\teff$ where
NH$_3$ has a very low abundance and of the condensate opacity that largely conceals the H$_2$ CIA
\footnote{This is no longer the case in low-metallicity L subdwarfs where H$_2$ CIA is noticeable 
\citep{burgasser03_sd,burgasser09,cushing09}.}.  We have thus focused on the role of these two
opacities in cloudless models, which are relevant to spectral types $\sim$ T4 and later.

\begin{figure}
\epsscale{0.85}
   \plotone{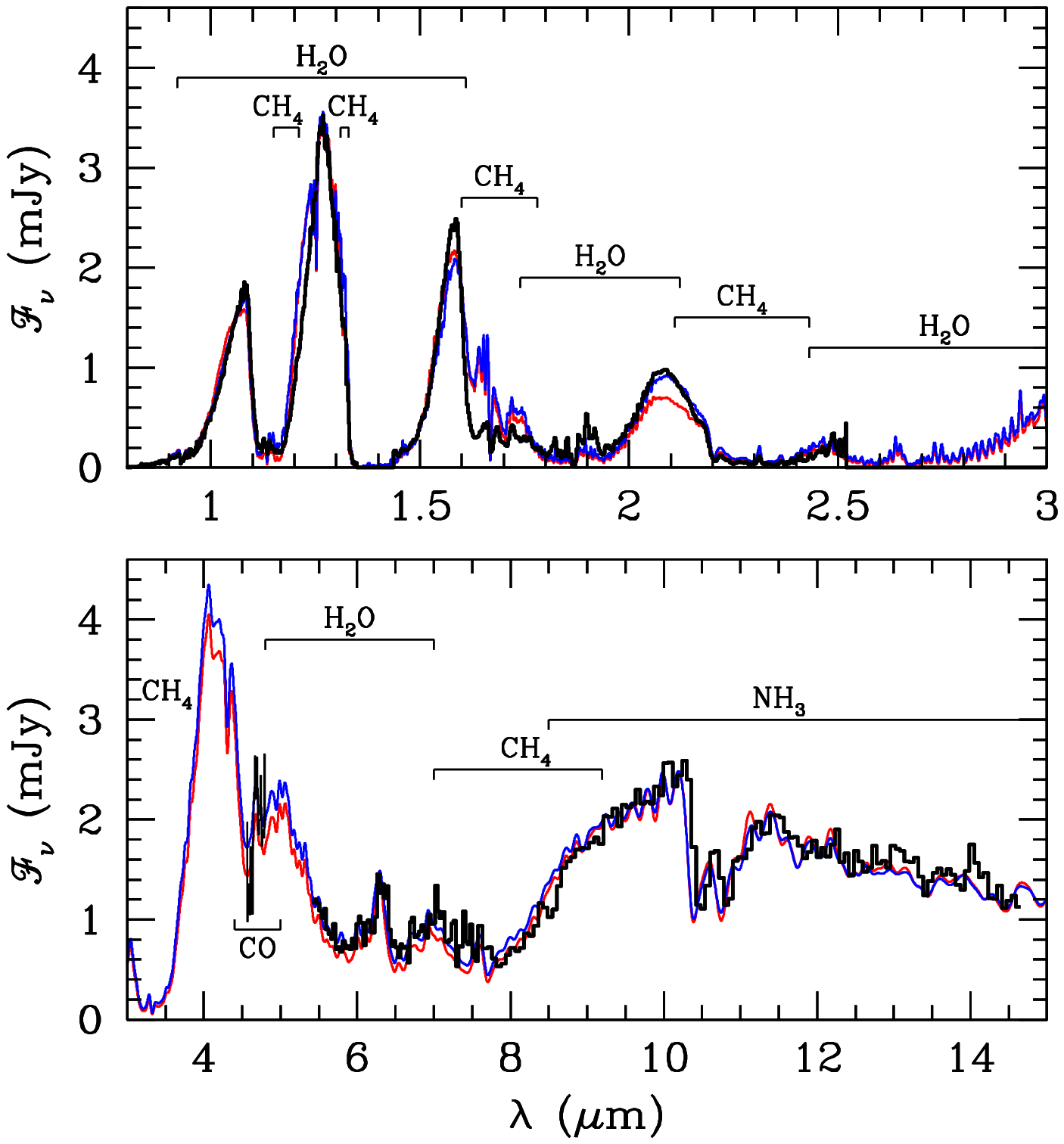}
   \caption{Spectrum of Gliese 570 D.  The data are shown in black \citep{burgasser03,geballe01,cushing06,geballe09}.
            The spectrum obtained with our
            previous models is shown in red \citep{saumon06}.  The blue curve shows the spectrum computed
            with the new H$_2$ CIA and the BYTe NH$_3$ line list with the same model parameters: $\teff=820\,$K,
            $\log g=5.23$ (cgs), [M/H]=0, and eddy diffusion coefficient $\log K_{zz}\,$(cm$^2$/s)=4.5. The models are plotted
            at a resolving power of $R=500$ (top panel) and $R=100$ (bottom panel), approximating that of
            the data. The models are not normalized to the data but are fluxes at Earth computed with the 
            known distance of Gl 570D and the radius obtained from the evolution with the above parameters.}
    \label{gl570d}
\end{figure}

\begin{figure}
\epsscale{0.70}
   \plotone{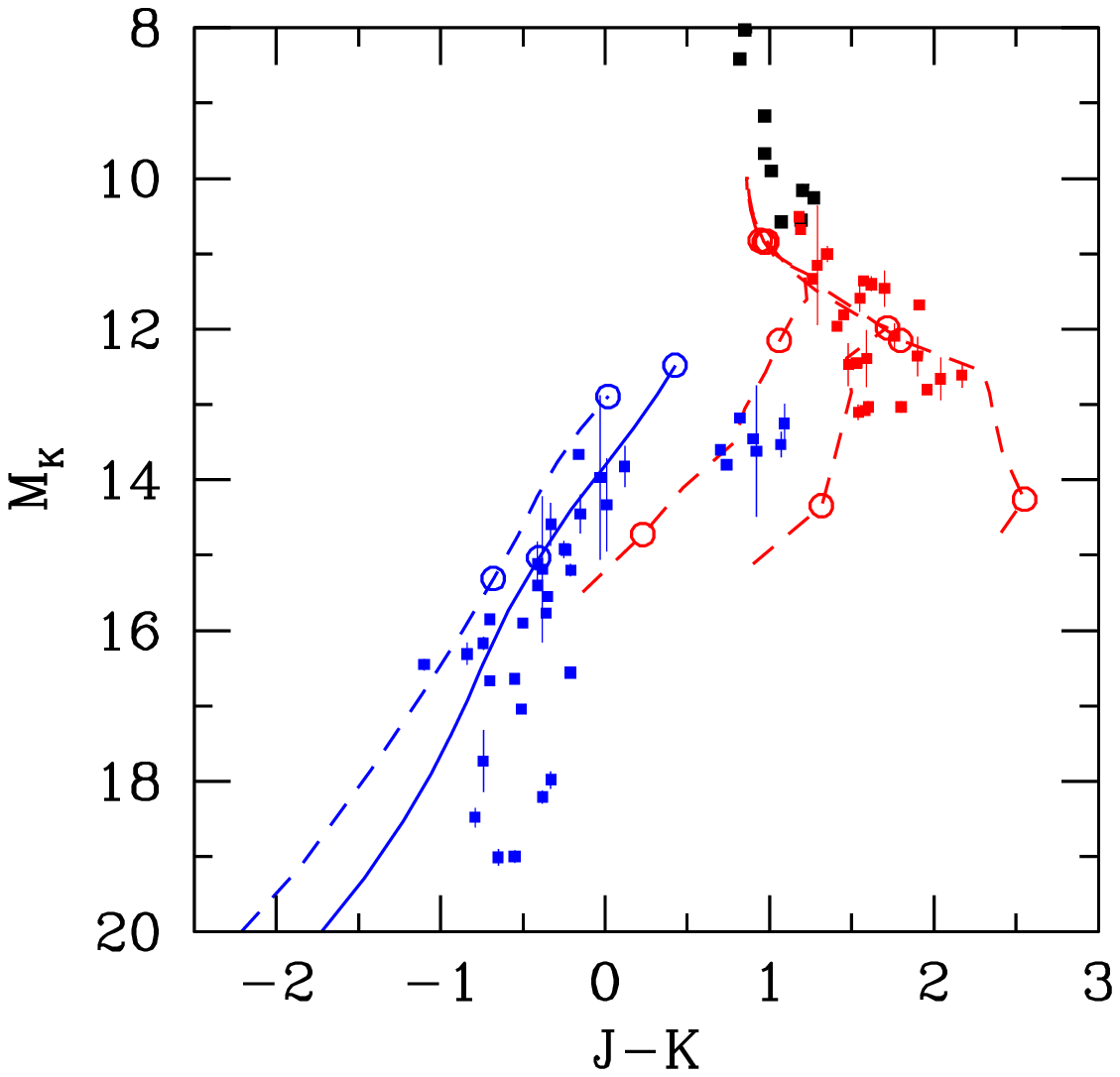}
   \caption{$M_K$ versus $J-K$ color-color diagram (MKO system) comparing models with the near infrared photometry of field MLT dwarfs.
            The photometry is from
            \citet{leggett02}, \citet{knapp04}, and \citet{marocco10} with M dwarfs in black, L dwarfs in red and T dwarfs in blue. All known binaries
            have been removed from the sample except those with resolved MKO photometry: 
            $\epsilon$ Indi B \citep{mccaugh04}, SDSS J102109.69$-$030420.1 and SDSS J042348.57$-$041403.5 \citep{burgasser06}, 
            and Kelu-1 \citep{ll05}.  The parallaxes are from \citet{perryman97}, \citet{dahn02}, \citet{tbk03}, \citet{vrba04},
            \citet{marocco10}, and various references in \citet{leggett02}. Solid blue curve: Cloudless models calculated with the
            new H$_2$ CIA and the BYTe NH$_3$ line list and an eddy diffusion coefficient
            $K_{zz}=10^4\,$cm$^2$/s (e.g. these model are not in chemical equilibrium).  The curve extends from $\teff=1500\,$K to 
            500$\,$K, with open circles indicating $\teff=1500\,$K and 1000$\,$K.  The dashed blue curve shows models
            computed with older H$_2$ CIA and NH$_3$ opacities and $K_{zz}=0$ (chemical equilibrium) \citep{sm08}.  
            The dashed red curves are cloudy models in chemical equilibrium from \citet{sm08} extending from 2400$\,$K to 900$\,$K with open
            circles indicating (from the top) $\teff=2000$, 1500, and 1000$\,$K.  Each curve corresponds to a different value of the
            cloud sedimentation parameter (see \citet{am01,sm08}) of $f_{\rm sed}=1$, 2 and 3, from right to left, respectively. Cloudy
            models computed with the new opacities overlap the cloudy sequences shown.
            All models shown have solar metallicity and $\log g=5$.} 
    \label{ccd1}
\end{figure}

\begin{figure}
\epsscale{0.75}
   \plotone{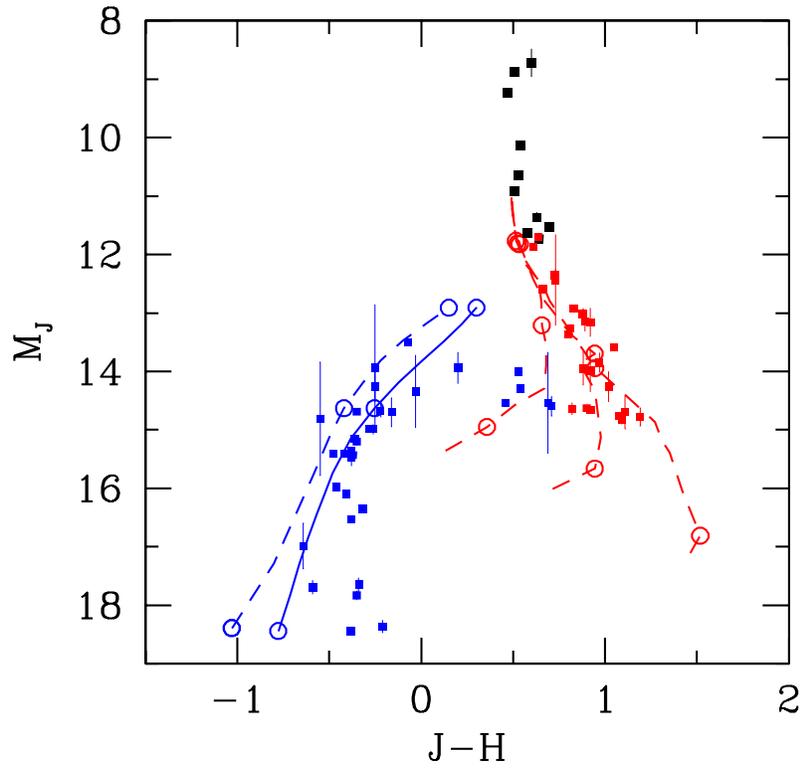}
   \caption{Same as Figure \ref{ccd1} for $M_J$ versus $J-H$. The Y0 dwarf WISEP J1541-2250, with $M_J=23.9\pm0.8$ and $J-H=0.17\pm0.63$ 
            \citep{cushing11,kirk11}, falls well outside this figure.}
    \label{ccd2}
\end{figure}

\begin{figure}
\epsscale{0.80}
   \plotone{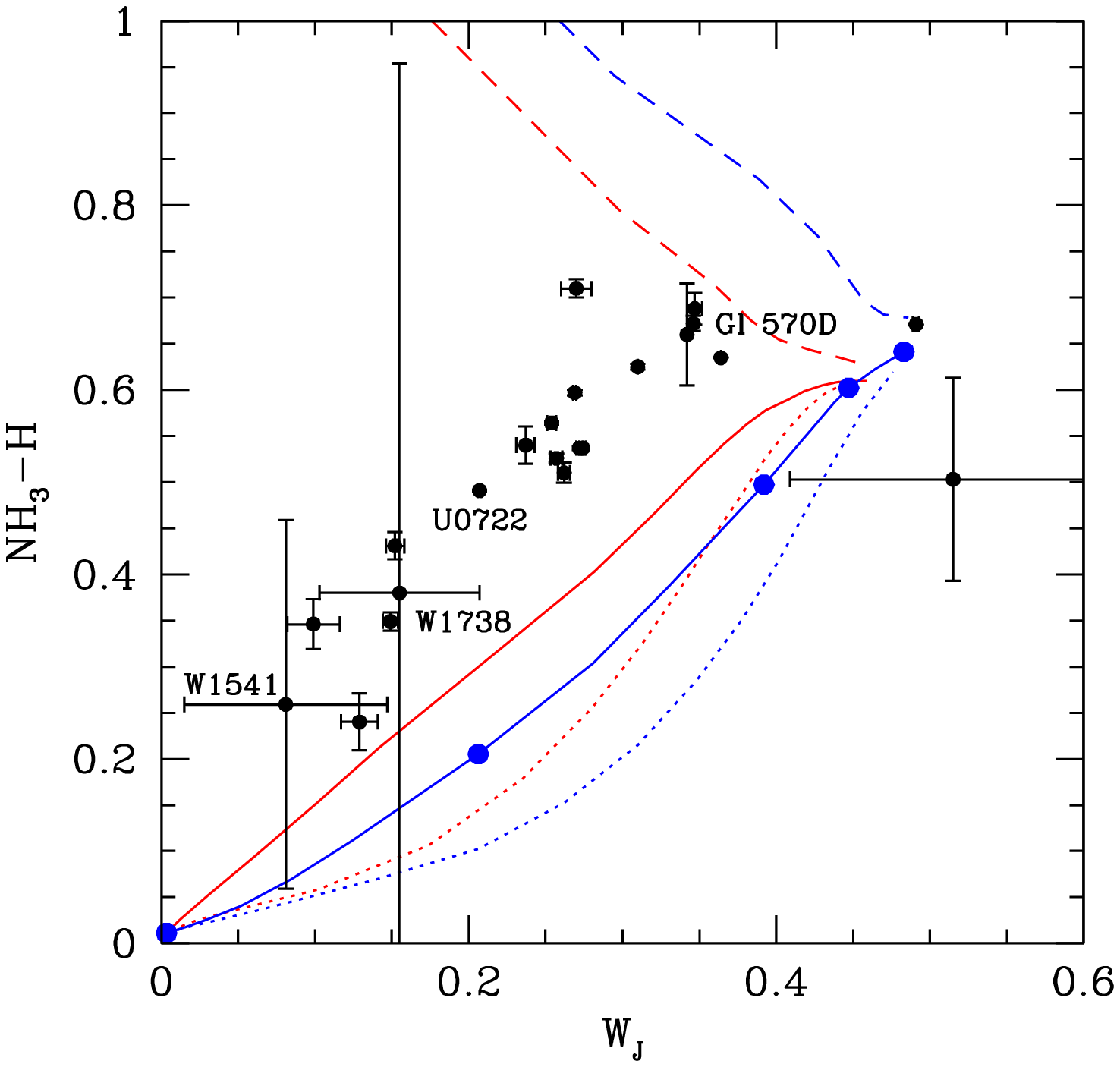}
   \caption{Spectral indices showing the growth of absorption in the blue wing of the $H$ band of late T dwarfs.
            The spectral indices NH$_3-H$ and $W_J$ are defined in \citet{delorme08} and \citet{warren07},
            respectively.  The curves show the indices computed from cloudless synthetic spectra starting at $\teff=200$ (lower 
            left corner) to 1000$\,$K (right) with solid dots marking steps of 200$\,$K. Two gravities,
            $\log g=4$ (cgs, red curves) and 5 (blue), are shown.  Dotted lines show models in chemical equilibrium and solid
            curves are models driven out of equilibrium by convective transport and mixing in
            the radiative zone parametrized by the eddy diffusion coefficient $K_{zz}=10^4\,$cm$^2$/s.
            The dashed lines show spectra computed with the same $(T,P)$ structures after removing the NH$_3$ opacity.
            The dwarfs Gl 570D (T8, \citet{delorme08}), UGPS 0722$-$05 (T9, \citet{cushing11}), WISEP J1738+2732 (Y0, \citet{cushing11},
            and WISEP J1541$-$2250 (Y0, \citet{cushing11}) are labeled.
            Data compiled from \citet{warren07,delorme08,burningham08,burningham09,burningham11a,burningham11b,cushing11}, and \citet{liu11}.}
    \label{nh3_index}
\end{figure}

\begin{figure}
\epsscale{1.00}
   \plotone{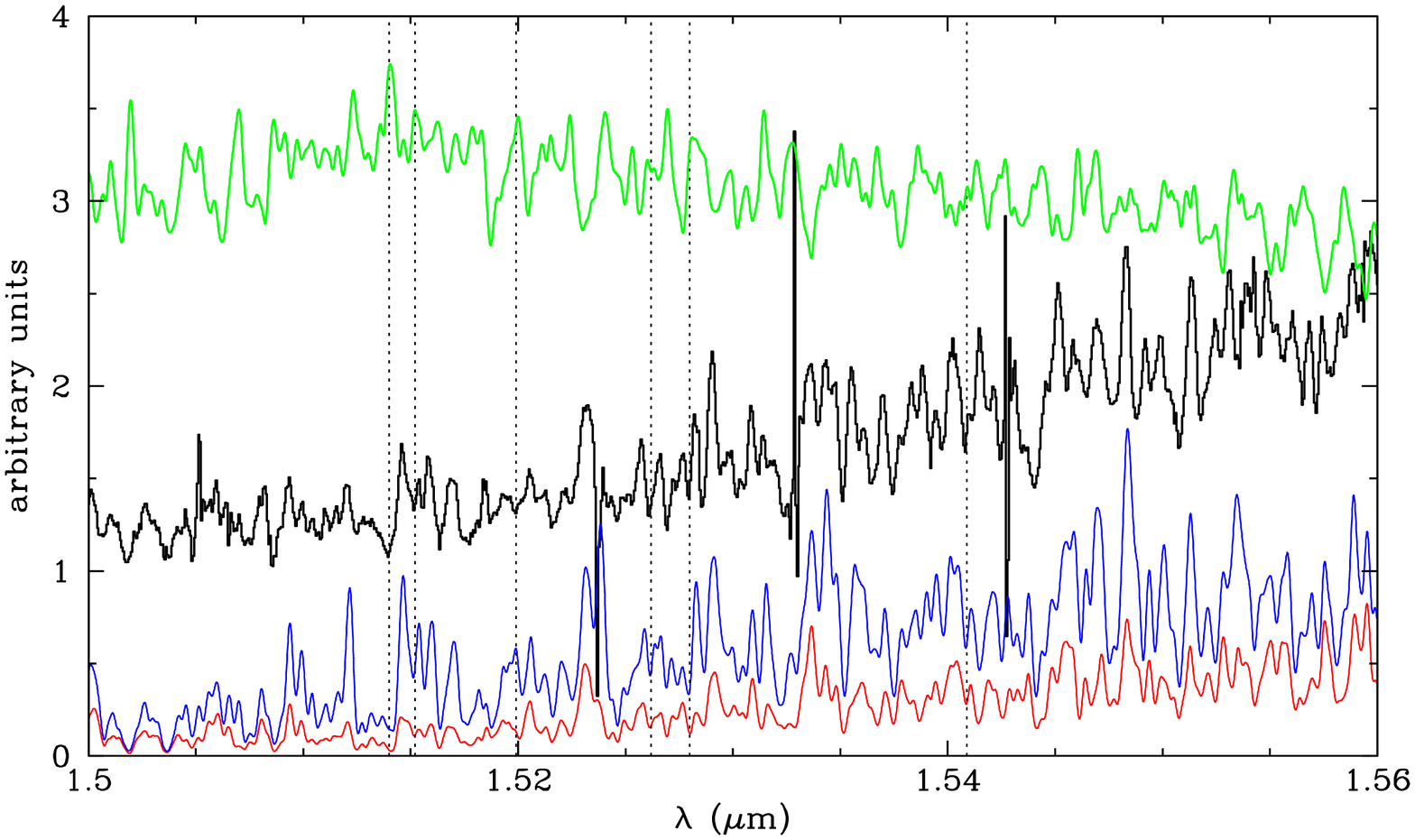}
   \caption{Medium resolution spectrum ($R\sim6000$) of UGPS0722$-$05 \citep{bochanski11} (black).
            The observed spectrum has been rescaled, shifted vertically, and the radial velocity (47$\,$km/s, \citet{bochanski11}) has been removed. 
            A cloudless synthetic spectrum, with $\teff=500\,$K, $\log g=4.25$, [M/H]=0, $K_{zz}=10^4\,$cm$^2$/s
            \citep{leggett11} is shown in red.  The blue curve represents the same spectrum
           calculated without any NH$_3$ opacity.  The $\log_{10}$ of the NH$_3$ opacity at 650$\,$K and 1 bar, 
           corresponding to the emission level of this part of the spectrum in the model, is shown in green at an arbitrary scale.
           The model spectra and the opacity are shown at the spectral resolution of the data.
           The vertical dotted lines show the NH$_3$ absorption features identified by \citet{bochanski11} in the 
           spectrum of UGPS 0722$-$05 which should correspond to peaks in the NH$_3$ opacity.}
    \label{0722_H}
\end{figure}

\begin{figure}
\epsscale{1.00}
   \plotone{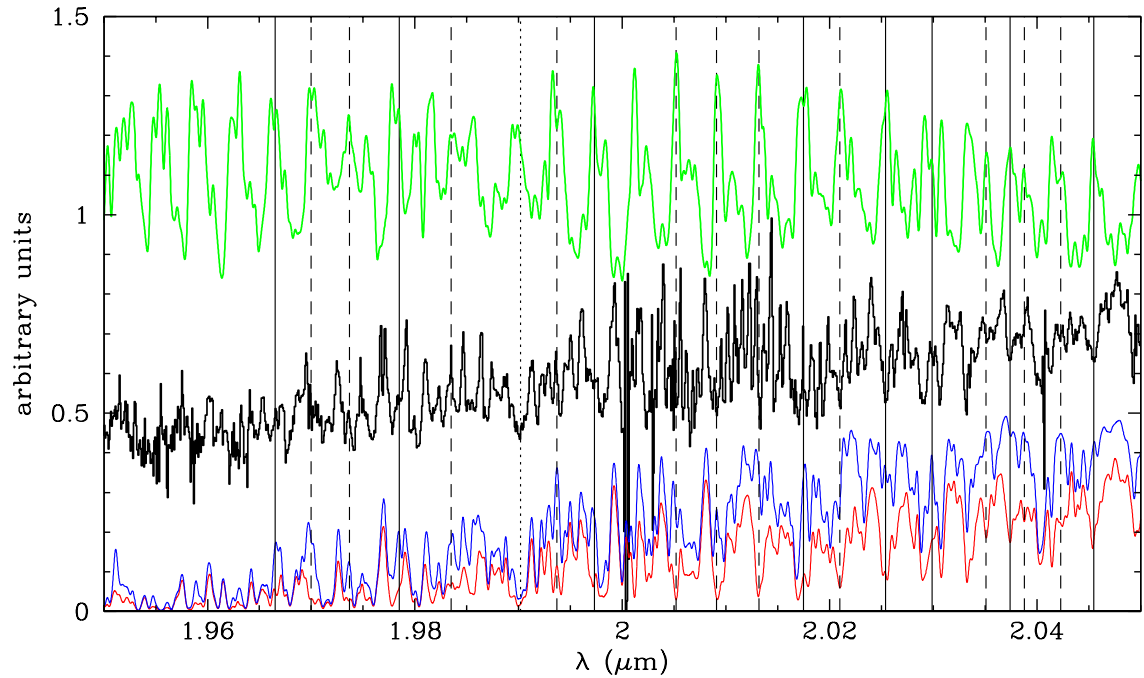}
   \caption{Same as Fig. \ref{0722_H} but in the $K$ band.  The models have been rescaled for clarity. 
            The vertical dotted line shows the NH$_3$ feature identified by \citet{bochanski11} in the spectrum of UGPS 0722$-$05.
            The other vertical lines are NH$_3$ absorption features predicted by the model spectrum, which all match peaks in the 
            NH$_3$ opacity.  Those that are also apparent in the data are shown with solid lines while those that are ambiguous or missing
            are shown with dashed lines.}
    \label{0722_K}
\end{figure}

The most notable change in the collision-induced absorption is a reduction of the opacity in the
first roto-vibrational band centered at 4200$\,$cm$^{-1}$ (2.4$\,\mu$m), resulting in an increased $K$
band flux. This corrects a deficiency in our earlier fit of the SED of Gl 570D \citep{saumon06}
and we now obtain a remarkable fit of nearly the entire SED of this late T dwarf (Fig. \ref{gl570d}).
Another consequence is that the cloudless sequence shifts toward bluer colors in 
near infrared color-magnitude diagrams and is in much better agreement with the colors of the late
T dwarfs in the field.  This new accurate H$_2$ CIA opacity will strengthen the analysis of gravity and
metallicity indicators in the spectra of late T dwarfs and L subdwarfs and help reduce the
uncomfortably large uncertainty in spectroscopic determinations of the gravity of brown dwarfs.

The new BYTe line list for NH$_3$ greatly improves the near infrared opacity of ammonia by adding 
several new bands
at short wavelengths as well as a very large number of ``hot transitions'' that fill in gaps
in the HITRAN line list. We can now reliably calculate the opacity of NH$_3$ at temperatures of
several hundreds of degrees, which is crucial even for the coolest known dwarfs as the $Y$, $J$ and 
$H$ bands typically probe
temperatures well above $\teff$.  The importance of this line list for the study of the T/Y transition
and the new Y dwarfs cannot be overstated.  Our models show that NH$_3$ develops very distinct signatures
in the $Y$ and $J$ band flux peaks, and confirms the tentative identification of NH$_3$ absorption
on the blue side of the $H$ band peak which had been anticipated by \citet{bsl03}.  More quantitatively, 
the new models reproduce the trend in the spectral indices NH$_3-H$ and $W_J$, confirming
that the rapid decrease in NH$_3-H$ at spectral types later than T8 is due to NH$_3$ absorption.
Thus, all the new WISE objects assigned to the Y dwarf class by \citet{cushing11} show NH$_3$
absorption in the $H$ band.  Furthermore, the data favors models where vertical transport reduces the
NH$_3$ abundance in the observable part of the atmosphere. At a more detailed level, the medium 
resolution near infrared spectrum
of the T9 UGPS 0722-05 \citep{bochanski11} provides a rare opportunity to compare data with the detailed predictions of
spectral features by the models and of the underlying molecular opacities.  While we found that none of 
the features pointed out by \citet{bochanski11} can be unambiguously attributed to NH$_3$, our new models 
allow the identification of 9 other NH$_3$ absorption features around 2$\,\mu$m.  The models clearly show that
while there is strong NH$_3$ absorption in the $H$ band in late T dwarfs and in the Y0 dwarfs, the
general behavior of the NH$_3$ absorption in the 1.55$\,\mu$m region makes it more difficult to
isolate individual features than in the 2.0$\,\mu$m region, requiring a resolving power of the
order of $R\sim 20000$ compared to $\sim 6000$ in the $K$ band.

Despite the progress that we have demonstrated here, obvious problems remain with the models. For instance, the fit of
Gl 570D still has discrepancies that stand out. The near infrared color-magnitude diagrams for field dwarfs
fail to reproduce the saturation (and possible turnover) revealed by the data at very low $\teff$. The spectral indices
NH$_3-H$ and $W_J$ do not yet provide a quantitative diagnostic of the atmospheric structure and chemistry.
With these two new calculations of molecular opacities that affect primarily late T dwarfs and the new class of Y dwarfs,
reliable opacities are available for all important gas phase absorbers except for CH$_4$. The inadequacy of the
the current line list of methane has been much maligned ever since the very first spectral modeling of a
T dwarf \citep{marley96,allard96}. We believe that for cloudless atmosphere models, the CH$_4$ opacity 
is the only remaining obstacle of significance to modeling the spectroscopic data accurately. 
The availability of a CH$_4$ line list comparable in quality and scope to the BYTe NH$_3$ line list would
open the door to reliable determinations of T dwarf gravities and metallicities, two fundamental parameters that
have remained elusive for the vast majority of brown dwarfs.

\acknowledgments

D.S., M.S.M., and R.S.F. acknowledge support from NASA Astrophysics Theory grant NNH11AQ54I.
L.F. and M.A. thank K.L.C. Hunt and her associates for the 
quantum chemical results provided to us prior to publication, which made their work possible. 
L.F. and M.A. also acknowledge NSF support through grants AST0708496 and AST0709106.

\bibliographystyle{apj}
\bibliography{references}

\clearpage


\clearpage

\clearpage

\clearpage

\clearpage

\clearpage

\clearpage

\clearpage

\clearpage

\clearpage

\clearpage

\clearpage

\clearpage

\clearpage

\clearpage

\clearpage
\end{document}